\def\year{2016}\relax
\documentclass[letterpaper]{article}
\usepackage{aaai16}
\usepackage{times}
\usepackage{helvet}
\usepackage{courier}
\usepackage{graphicx}
\usepackage{caption}
\usepackage{subcaption}
\usepackage{url}
\usepackage{amsmath}
\usepackage{amsfonts}
\usepackage{multirow}
\usepackage{amsmath}

\usepackage[table,xcdraw]{xcolor}
\usepackage[linesnumbered,lined,boxed,commentsnumbered, ruled]{algorithm2e}
\SetAlFnt{\small}
\newcommand{\newcite}[1]{\citeauthor{#1}~\shortcite{#1}}
\usepackage[english]{babel}

\usepackage{enumerate}

\begin{document}
% \title{Content Speaks For Itself: Diversity Link based Network Learning Method}
\title{A General Framework for Content-enhanced Network Representation Learning}
\author{Xiaofei Sun, Jiang Guo, Xiao Ding \and Ting Liu \\
Center for Social Computing and Information Retrieval, Harbin Institute of Technology, China \\
{\tt \{xfsun, jguo, xding, tliu\}@ir.hit.edu.cn}\\
}
\maketitle

% !TEX root = main.tex
\begin{abstract}
    This paper investigates the problem of network embedding, which aims at learning low-dimensional vector representation of nodes in networks.
    % , is useful for various tasks like node classification and link prediction.
    Most existing network embedding methods rely solely on the network structure, i.e., the linkage relationships between nodes, but ignore the rich content information associated with it, which is common in real world networks and beneficial to describing the characteristics of a node.
    In this paper, we propose content-enhanced network embedding (\textsc{CENE}), which is capable of jointly leveraging the network structure and the content information.
    Our approach integrates text modeling and structure modeling in a general framework by treating the content information as a special kind of node.
    Experiments on several real world networks with application to node classification show that our models outperform all existing network embedding methods, demonstrating the merits of content information and joint learning.
    % Also note that the presented approach can be flexibly scale to other modalities of content such as images and speech, and thus is more practically useful.
\end{abstract}

% To show the merits of content information, we propose a novel network embedding approach named CENE, that can jointly leverage the network structure and the content information (texts) associated with it.
% We focus on the textual contents in this study, but the presented approach can be flexibly scale to various modalities of contents such as images and speech.

% !TEX root = main.tex

\section{Introduction}
\label{sec:introduction}

% * Network structure is common and useful

Network embedding, which aims at learning low-dimensional vector representations of a network, has attracted increasing interest in recent years. It has been shown highly effective in many important tasks in network analysis involving predictions over nodes and edges, such as node classification \cite{tsoumakas2006multi,sen2008collective}, recommendation \cite{tu2014inferring,yu2014personalized} and link prediction \cite{liben2007link}.
% Recent efforts have been made to overcome this problem by learning a latent space representation for the entities in a network, such as Deepwalk\cite{perozzi2014deepwalk}, LINE \cite{tang2015line}, GraRep \cite{cao2015grarep}, and node2vec \cite{grovernode2vec}.

Various approaches have been proposed toward this goal, typically including \textit{Deepwalk} \cite{perozzi2014deepwalk}, \textit{LINE} \cite{tang2015line}, \textit{GraRep} \cite{cao2015grarep}, and \textit{node2vec} \cite{grovernode2vec}.
% * Former work is good, but is limited.
These models have been proven effective in several real world networks.
Most of the previous approaches utilize information only from the network structure, i.e., the linkage relationships between nodes, while paying scant attention to the content of each node, which is common in real-world networks.
In a typical social network with users as vertices, the user-generated contents (e.g., texts, images) will serve as rich extra information which should be important for node representation and beneficial to downstream applications.
% These kind of information has been proved to be effective for various kind of tasks based on network entities.

\begin{figure}
    \centering
    \includegraphics[width=80mm]{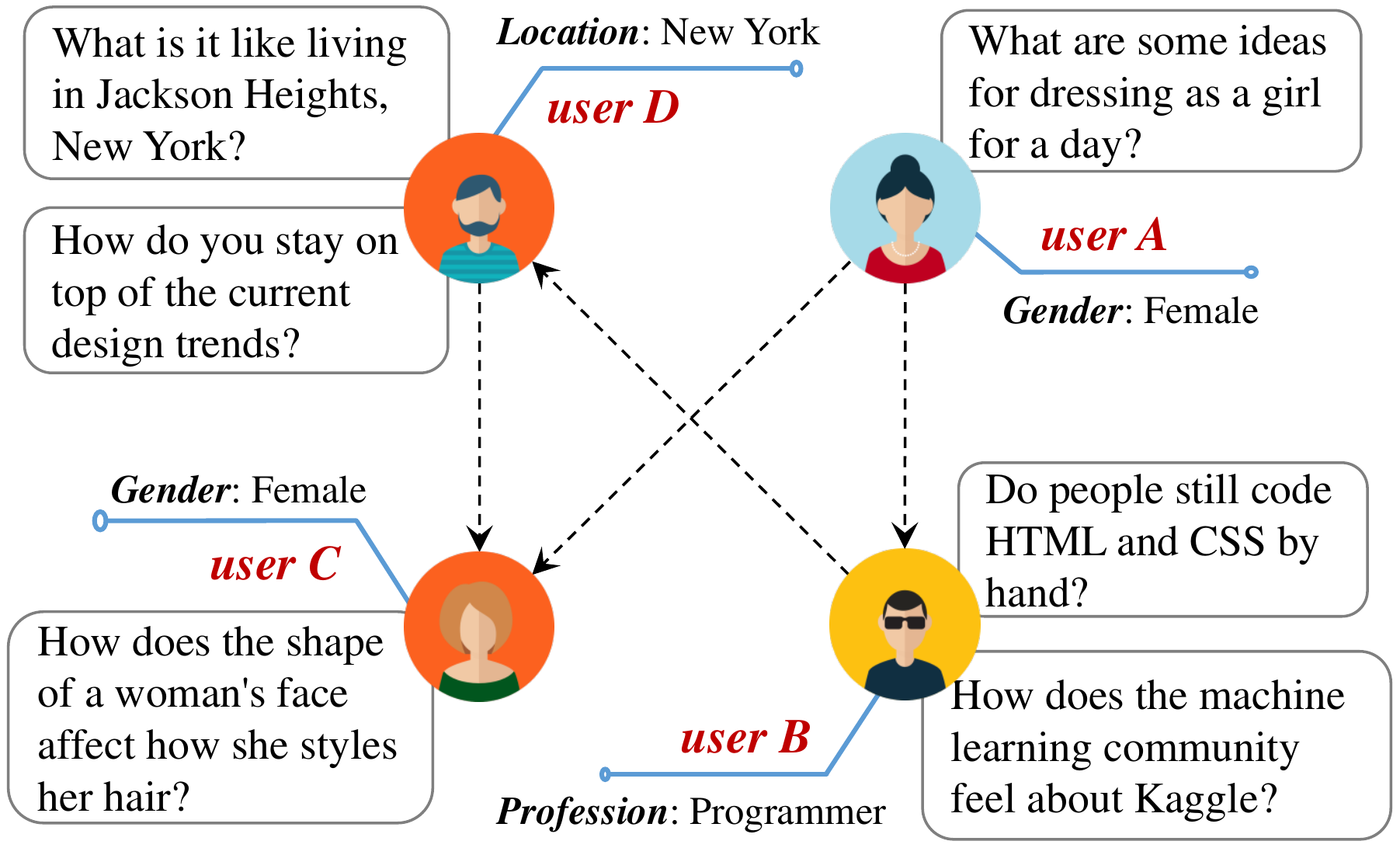}
    \caption{
    A toy network of Quora users with the content being titles of questions that user has followed.
    }
    \label{fig:example}
    % \vspace{-0.8em}
\end{figure}

Figure \ref{fig:example} shows an example network from Quora, a community question answering website.
Users in Quora can \textit{follow} each other, creating directed connections in the network.
More importantly, users are expected to ask or answer questions, which can be treated as users' contents.
These contents are critical for identifying the characteristics of users, and thus will significantly benefit tasks like node classification (e.g. \textit{gender}, \textit{location} and \textit{profession}).
For example, we can infer from the contents of \emph{user A} and \emph{user C} (Figure \ref{fig:example}) that they are likely to be female users (\textit{gender}).
Besides, \emph{user B} is supposed to be a programmer (\textit{profession}) and \emph{user D} probably lives in New York (\textit{location}).

% * An intuitive solution

% To cope with this challenge, few studies like TADW \newcite{yang2015network} and TriDNR \newcite{pan16tri} have taken text information into account, which inspire our work.
To cope with this challenge, \newcite{yang2015network} presented text-associated DeepWalk (TADW), which incorporates textual features into network embeddings through matrix factorization. This approach typically suffers from high computational cost and not scalable to large-scale networks. Besides, contents in TADW are simply incorporated as unordered text features instead of being explicitly modeled. Therefore, deeper semantics contained in the contents cannot be well captured.

% More recently, \newcite{pan16tri} described a semi-supervised model (TriDNR) which exploits inter-node, node-word, as well as label-word relationship for network representation learning.

% * Our method to solve this problem

\textbf{Present work.} We present a general framework for learning Content-Enhanced Network Embedding (CENE) that is capable of jointly leveraging the network structure and the contents.
We consider textual contents in this study, however, our approach can be flexibly scaled to other modalities of content.
Specifically, each piece of content information (e.g., a tweet one posts in twitter, a question one follows in Quora) is formalized as a \textit{document}, and we integrate each document into the network by creating a special kind of node, whose representation will be computed compositionally from words.
The resulting augmented network will consist of two kinds of links: the \textit{node-node} link and the \textit{node-content} link.
By optimizing the joint objective, the knowledge contained in the contents will be effectively distilled into node embeddings.

% * Summarize

To summarize, we make the following contributions:

\begin{itemize}
    % \vspace{-0.3em}
    \item We propose a novel network embedding model that captures both textual contents and network structure. Experiments on the tasks of node classification using two real world datasets demonstrate its superiority over various baseline methods.% \vspace{-0.3em}
          % \item We compare three sentence embedding methods in our model and analyze their (dis)advantages.\vspace{-0.3em}
    \item We collect a network dataset which contains node attributes and rich textual contents. It will be made publicly available for research purpose. % \vspace{-0.3em}
\end{itemize}

% !TEX root = main.tex

\section{Related Work}

\subsection{Text Embedding}

In order to obtain text embeddings (e.g., sentence, paragraph), a simple and intuitive approach would be averaging the embeddings of each word in the text \cite{mitchell2010composition,ferrone2013linear,iyyer2015deep}. More sophisticated models have been designed to utilize the internal structure of sentences or documents to assist the composition. For example, \newcite{socher2013recursive} and \newcite{socher2014grounded} use recursive neural networks over parse trees to obtain sentence representations. To alleviate the dependency on syntatic parsing, convolutional neural networks (CNN) \cite{blunsom2014convolutional,DBLP:conf/naacl/Johnson015} are employed which use simple bottom-up hierarchical structures for composition. Another alternative model is the LSTM-based recurrent neural network (RNN) \cite{kiros2015skip}, which is a variant of RNN that uses long short-term memory cells for capturing long-term dependencies.
% It utilizes the sequential structure, however, recent work has demonstrated that LSTM can implicitly captures  tree structure \cite{bowman2015tree}.

\subsection{Network Embedding}

\newcite{hoff2002latent} first propose to learn latent space representation of vertices in a network. Some earlier works focus on the feature vectors and the leading eigenvectors are regarded as the network representations, e.g., MDS \cite{borg2005modern}, IsoMap \cite{tenenbaum2000global}, LLE \cite{roweis2000nonlinear}, and Laplacian Eigenmaps \cite{belkin2001laplacian}.

Recent advancements include DeepWalk \cite{perozzi2014deepwalk}, which learns vertex embeddings using the skip-gram model \cite{mikolov2013distributed} on vertex sequences generated by random walking on the network.
Inspired by Deepwalk, walklet \cite{perozzi2016walklets} focuses on multiscale representation learning, node2vec \cite{grovernode2vec} explores different random walk strategies and \newcite{DBLP:conf/kdd/OuCPZ016} emphasises the asymmetric transitivity of a network.
Some others focus on depicting the distance between vertices. LINE \cite{tang2015line} exploits both first-order and second-order proximity in an network while \newcite{cao2015grarep} expand the proximity into $k$-order (or $k$-step) and integrates global structural information of the network into the learning process. These methods could also be applied to prediction tasks in heterogeneous text networks \cite{tang2015pte}. Another attempt is based on the factorization of relationship matrix \cite{yang2015comprehend}. Most recently, \newcite{wangstructural} adopt a deep model to capture the non-linear network structure.

\newcite{yang2015network} present the first work that combines structure and content information for learning network embeddings. They show that DeepWalk is equivalent to matrix factorization (MF) and text features of vertices can be incorporated via factorizing a text-associated matrix.
This method, however, suffers from the high computation cost of MF and has difficulties scaling to large-scale networks.
% However, matrix fraction is very computationally expensive for large-scale network data. And the situation will be worse when some vertices don't have contents.
\newcite{pan16tri} instead combines DeepWalk with Doc2Vec \cite{le2014distributed}, along with partial labels of nodes that constitutes a semi-supervised model. However, Doc2Vec is far from being expressive of the contents. Besides, it cannot generalize to other modalities of contents like images.

% They However, their model effects from word level, which means it ignore the semantic information, which is important for document embedding, and it's also hard to expand to other content forms like images.

% Both of TADW and TriDNR suppose each node only have one content, which is unreasonable for some real world networks, especially social networks. More over, as their approaches put too many attention on word-level informations, it's hard to apply these methods to other kinds of contents such as images and videos.

% !TEX root = main.tex

\section{Problem Definition}

\textbf{Definition 1. (Network)}
Let $G=(V,E,C)$ denote a network, where $V$ is the set of vertices, representing the nodes of the network; $E \subseteq (V \times V)$ is the set of edges, representing the relations between the nodes; and $C$ denotes the contents of nodes. $C=\{(v,doc)| v \in V;doc=\{S_i\}\}$, where $S_i$ denotes $i$-th sentence of $doc$ and is composed of word sequence $<w_1,w_2, \dots ,w_n>$.
Without loss of generality, we assume the structure of network to be a directed graph.\footnote{Undirected networks can be readily converted to directed ones by replacing each undirected edge with two oppositely directed edges.}

\noindent\textbf{Definition 2. (Network Embedding)}
Given a network denoted as $G=(V,E,C)$, the aim of network embedding is to allocate a low dimensional real-valued vector representation $e_v \in \mathbb{R}^d$ for each vertex $v \in V$, where $d  \ll |V|$. Let $\theta= (e_1,e_2, \dots  ,e_{|V|})$ denotes the embedded vectors in the latent space.
$\theta$ is supposed to maintain as much topological information of the original network as possible.

As $e_v$ can be regarded as a feature vector of vertex $v$, it is straightforward to use it as the input of subsequent tasks like node classification. Another notable trait is that this kind of embedding is not task-specific so that it can be applied to different kinds of tasks without retraining.

% !TEX root = main.tex

\section{Method}

\subsection{General Framework}

To maintain the structural information of a network, we describe a general framework that minimizes the following objective:
\begin{equation}
    \begin{aligned}
        \mathcal{L}_g = & \sum_{(u,v) \in SP}{\log{p(u,v;\theta)}}+\sum_{(u,v) \in SN}{\log{(1-p(u,v;\theta))}} \\
    \end{aligned}
    \label{eq:general_loss}
\end{equation}
where $SP$ is the set of positive vertex pairs and $SN$ is negative pair set.
For example, in random walk-based algorithms (Deepwalk, walklet, node2vec), $SP$ is the set of adjacent vertex pairs in the routes generated through random walking, and $SN$ is the union of all negative sampling sets.
$p(u,v;\theta)$ is the joint probability between vertex $u$ and $v$, which means the probability of pair $(u,v)$ existing in $SP$ and correspondingly $1-p(u,v;\theta)$ is the probability that $(u,v)$ does not exist.
% Correspondingly, $S_p$ is the first and second order vertex pair and $S_n$ is union of pairs not in first order pair sets and negative sampling results for second order pairs.

To further utilize the content information, a simple way is to concatenate the content embedding with the node embedding, both of which are trained independently. Formally, let $e_u=e_u \oplus f_e(C_u)$ be the representation of node $u$, where $C_u=\{(u,c)|(u,c) \in C\}$ is the set of all contents of node $u$. This method, however, requires each node in the network to be associated with some contents, which is too rigid for real world networks. % weak adaptability for multiple content and nodes without contents.
% , which makes it hard to be applied to real world networks.

\begin{figure}[t]
    \centering
    \includegraphics[width=80mm]{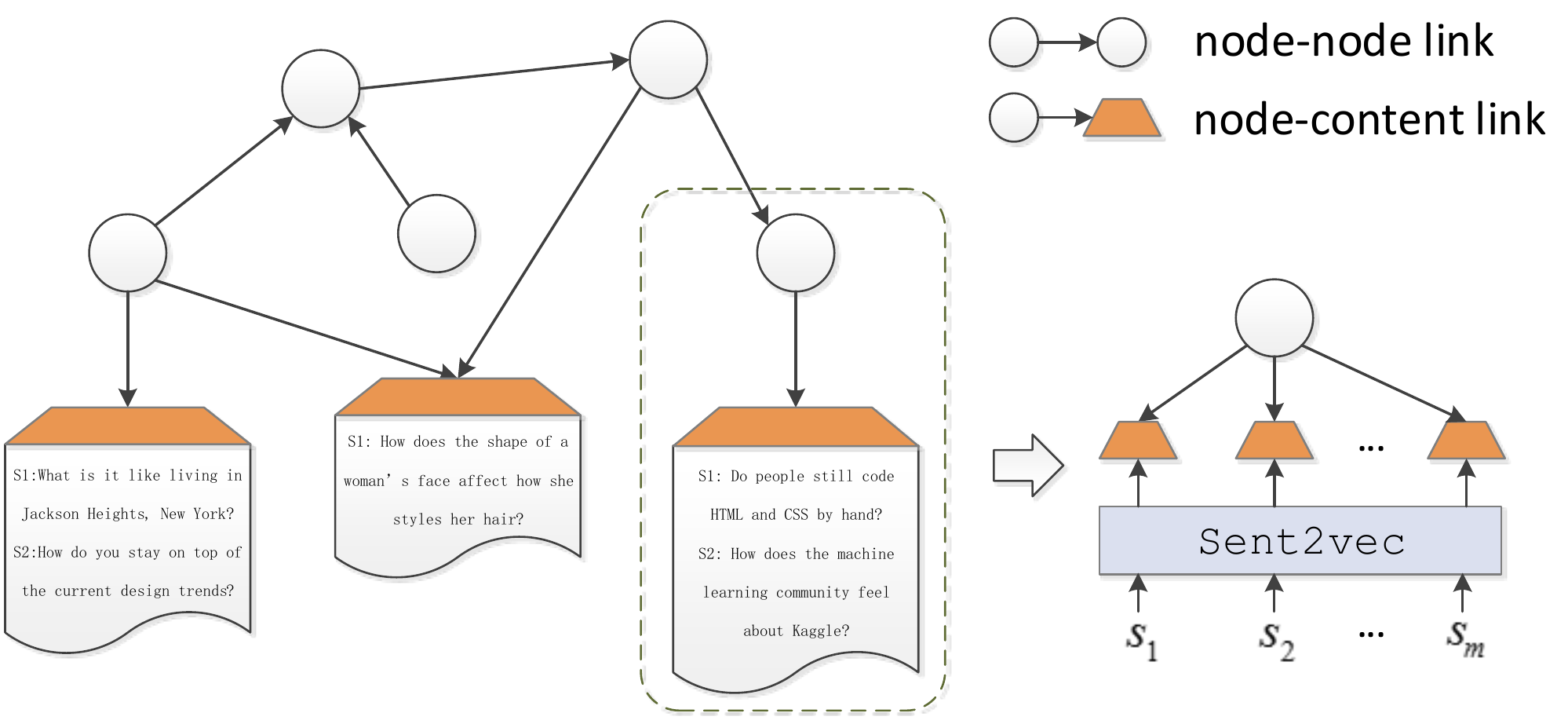}
    \caption{Illustration of our framework.}
    \label{fig:framework}
    % \vspace{-1.0em}
\end{figure}

In this paper, we introduce contents (documents) as a special kind of nodes, and then the augmented network can be represented as: $G^{aug}=(V_n, V_c, E_{nn}, E_{nc})$, where $V_n=V$ is vertex set; $V_c=\{c|(u,c) \in C, u \in V\}$ is the content set; $E_{nn}=E$ is the set of edges between vertices; and $E_{nc}=V_n \times V_c$ is the set of edges between vertices and contents. In this way, different nodes can also interact through connection with a same content node (e.g., two Twitter users retweet the same post), which significantly alleviates the structural sparsity in $E_{nn}$.
% Note that $E_{nc}$ can be either one-to-many or many-to-many. That is, different nodes can share parts of same contents. For example, different twitter users may retweet the same post.
The resulting framework structure is illustrated in Figure~\ref{fig:framework}.

% With the above input $G$ and according to the Eq.\ref{eq:general_loss}, we could construct loss function for both node-node links and node-content links. By combining these two losses we can finally get the objective function.
Next, we will describe the loss functions involving node-node links and node-content links respectively, following the notation in Eq.\ref{eq:general_loss}.

\subsection{Node-Node Link}

For node-node links, we specify $SP$ as $E_{nn}$. Inspired by the idea of negative sampling~\cite{mikolov2013distributed}, we sample a set $SN_{nn}^u=\{v^\prime|(u,v^\prime) \notin E_{nn}\}$ for each edge $(u,v)$. Then $SN = \underset{u \in V_{nn}}{\cup}SN_{nn}^u$.

\begin{equation}
    \mathcal{L}_{nn} = \sum_{(u,v) \in E_{nn}}{ [\log{p(v,u;\theta)}-\sum_{v^\prime \in SN_{nn}^u}{\log{p(v^\prime,u;\theta)}}] }
    \label{eq:nn_loss}
\end{equation}

% The critical problem is how to calculate $p(v,u)$. Here we follow the Ski
Here $p(v,u)$ (we omit $\theta$ for simplicity) is computed using a logistic function:
\begin{equation}
    p(v,u)=\sigma(e_u \cdot e_v) = \frac{1}{1 + \exp (-e_u \cdot e_v)}
    \label{eq:p_u_v_sy}
\end{equation}
% and $\sigma(\cdot)$ is logistic function.

However, Eq.\ref{eq:p_u_v_sy} is a symmetrical operation, which means $p(v,u)=p(u,v)$, and this is not suitable for directed networks.
So we splited $\forall e_u=(e_u^{in}, e_u^{out})$ where $e_u^{in} \in \mathbb{R}^{d/2}$ and $e_u^{out} \in \mathbb{R}^{d/2}$.
Then $p(v,u)$ can be computed as:
\begin{equation}
    p(v,u)=\sigma(e_v^{in} \cdot e_u^{out}) = \frac{1}{1 + \exp (-e_v^{in} \cdot e_u^{out})}
    \label{eq:p_u_v_asy}
\end{equation}

\subsection{Node-Content Link}

\begin{figure}[t]
    \centering
    \includegraphics[width=70mm]{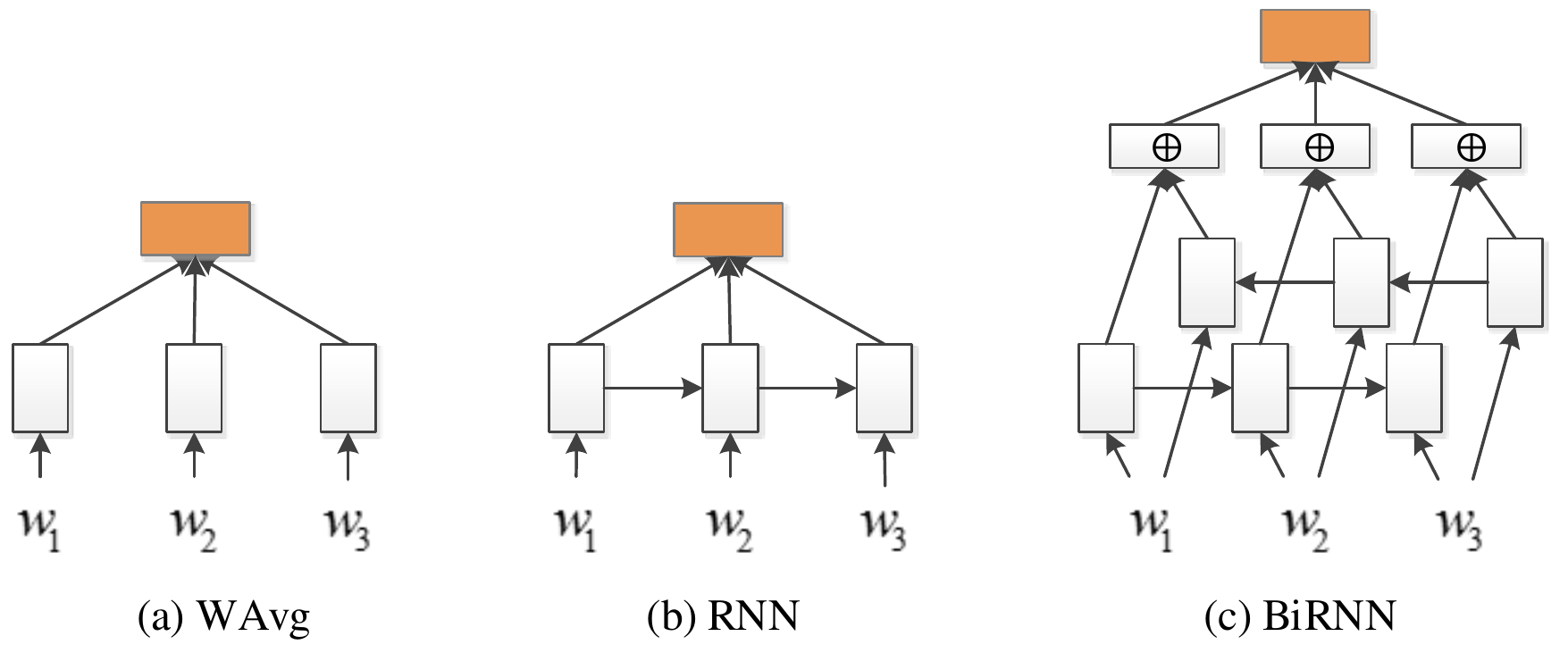}
    \caption{Sentence modeling approaches.}
    \label{fig:sent2vec}
    % \vspace{-0.8em}
\end{figure}

The node-content loss is similar to Eq.\ref{eq:nn_loss}. Let  $SN_{nc}^u=\{c^\prime|(u,c^\prime) \notin E_{nc}\}$ denote the negative sampling set for edge $(u,c)$, then the loss can be written as:
\begin{equation}
    \mathcal{L}_{nc} = \sum_{(u,c) \in E_{nc}}{ [\log{p(c,u;\theta)}-\sum_{c^\prime \in SN_{nc}^u}{\log{p(c,u;\theta)}}] }
    \label{eq:nc_loss}
\end{equation}
where
\begin{equation}
    p(c,u)=\sigma(e_u,f_e(c))=\sigma(e_u^{out} \oplus e_u^{in},f_e(c))
    \label{eq:p_u_c}
\end{equation}
Instead of allocating an arbitrary embedding for each document $c$, here, we use a composition function $f_e(\cdot)$ to compute the content representation in order to fully capture the semantics of texts.
In this paper, we further decompose each document into sentences, and model \textit{node-sentence link} separately (Figure~\ref{fig:framework}).
We investigate three typical composition models for learning sentence representations (Figure~\ref{fig:sent2vec}).

% Different with node embedding, it's hard to allocate a arbitrory unque embedding for each content, because is composed of sequences of words and can be unlimited different sentences, not to mention contents.

% Therefor, we treat content embedding as an output of embedding fuction $f_e(\cdot)$. In this paper, we have specify the content as text content and explore three different typical methods for document embedding (Figure~\ref{fig:sent2vec}).

\textbf{Word Embedding Average (WAvg)}. This approach simply takes the average of word vectors as the sentence embedding. Despite its obliviousness to word order, it has proved surprisingly effective in text categorization tasks~\cite{joulin2016bag}.
\begin{equation}
    f_e(c)=\frac{1}{|c|}\sum_{i}e_{w_i}
\end{equation}
% % CBOW has proven useful in different tasks and is easy to compute, making it an important model class to consider.

\textbf{Recurrent Neural Network (RNN)}. Here we use the gated recurrent unit (GRU) proposed by \newcite{cho2014learning}. GRU is a simplified version of the LSTM unit proposed earlier \cite{hochreiter1997long}, with fewer parameters while still preserving the ability of capturing long-term dependencies. Instead of simply using the hidden representation at the final state as the sentence representation, we apply mean pooling over all history hidden states:
\begin{equation}
    \overrightarrow{h_i} = \mathtt{GRU}(e_{w_i}, h_{i-1});~~~~
    f_e(c) = \frac{1}{|c|}\sum_{i}\overrightarrow{h_i}
\end{equation}

\textbf{Bidirectional Recurrent Neural Network (BiRNN)}. In practice, even with GRU, RNN still cannot capture very long-term dependencies well. Hence, we further adopt a bidirectional variant \cite{schuster1997bidirectional} that processes a sentence in both directions with two separate hidden layers. The hidden state vectors from two directions' GRU units at each position are then concatenated, and finally passed through a mean pooling layer.
\begin{equation}
    \overleftarrow{h_i} = \mathtt{GRU}(e_{w_i}, h_{i+1});~~~~
    f_e(c) = \frac{1}{|c|}\sum_{i}(\overrightarrow{h_i}\oplus \overleftarrow{h_i})
\end{equation}

\subsection{Joint Learning}

% With the two objective function in Eq.\ref{eq:nn_loss} and Eq.\ref{eq:nc_loss}, we optimize the weighted combination of them. Specifically, we aim at mimizing the following objective function:
Finally, we optimize the following joint objective function, which is a weighted combination of the node-node loss (Eq.\ref{eq:nn_loss}) and the node-content loss (Eq.\ref{eq:nc_loss}):
\begin{equation}
    \mathcal{L} = \alpha*\mathcal{L}_{nn} +(1-\alpha)*\mathcal{L}_{nc}
    \label{eq:loss}
\end{equation}
where $\alpha \in [0,1]$ is a parameter to balance the importances of the two objectives.
With the $\alpha$ increasing, more structure information (node-node link) will be taken into consideration.
All parameters, including $\theta$ and parameters in $f_e(\cdot)$ are jointly optimized.

We use stochastic gradient descent (SGD) with learning rate decay for optimization.
The gradients are computed with back-propagation.
In our implementation, we approximate the effect of $\alpha$ through instance sampling (node-node and node-content) in each training epoch.
More details are shown in Algorithm \ref{al:joint_training}.

% !TEX root = main.tex

% \SetAlFnt{\footnotesize}
% \vspace{-0.7em}
\begin{algorithm}
    \caption{Joint Training.}
    \label{al:joint_training}
    \SetKwData{Index}{Index}
    % \begin{scriptsize}
    \textbf{Input}: $V_n$, $V_c$, $E_{nn}$, $E_{nc}$, balance weight $\alpha$, content embedding method $f_e(\cdot)$, $\eta$, $MaxStep$

    \For{ $step \leftarrow 0$ \KwTo $MaxStep$ }{
    Random generate $x \sim N(0,1)$\;
    \eIf{$x<\alpha$}{
    Get negative sampling set $SN_{nn}^u$\;
    random select $(u,v)$ from $E_{nn}$\;
    Lookup embedding of $u$ and $v$ from $\theta$\;
    Lookup embedding of $v^\prime \in SN_{nn}^u$ from $\theta$\;
    }{
    Get negative sampling set $SN_{nc}^u$\;
    Lookup embedding of $u$ from $\theta$\;
    Calculate embedding of $c$ and $\forall c^\prime \in SN_{nc}^u$ with $f_e$\;
    }
    Perform SGD on the corresponding loss;
    $\eta \leftarrow \frac{\eta}{1-\frac{step}{MaxStep}}$\;
    }
    % \end{scriptsize}
\end{algorithm}

% !TEX root = main.tex

\section{Experiments Setup}

\subsection{Dataset}

We conduct experiments on two real world datasets: DBLP \cite{Tang:08KDD} and Zhihu. An overview of these networks is given in Table~\ref{tab:network_overview}.

\subsubsection{DBLP}
We use the DBLP\footnote{\url{cn.aminer.org/citation} (V1 is used)} dataset to construct the citation network.
Two popular conferences: SIGIR and KDD, are chosen as the two categories for node classification.\footnote{Note that SIGIR mainly focuses on information retrieval while KDD is more related to data mining.}
% Papers published in the conferences are assumed to belong to the categories corresponding to the conferences.
Here each paper is regarded as a node, and every directed edge between two nodes indicate a citation.
We use the abstract of each paper as the contents.
Note that only 16.7\% nodes on DBLP have contents and we keep all nodes of DBLP for experiments.
% The content of DBLP is the abstract of each paper.
% For RNN is weak to get a good embedding for long sententce, we split each abstract into sentences and treat each sentence as a single content.

\subsubsection{Zhihu}
Zhihu\footnote{\url{www.zhihu.com}} is a Chinese community social-network based Q\&A site, which aims at building a knowledge repository of questions and answers created and organized by users.
% The contents of the Zhihu network are the question titles that each user follows.
We first collected the users' following lists, following questions list and their profiles.
Then, we construct the Zhihu network with users as vertices, and edges indicating the following relationships.
The question titles that each user follows are used as their associated contents.
% The vertices on Zhihu networks represent users and the edges indicate the following relationship.

We select the top three frequent attributes for our experiments: \textit{gender}, \textit{location} and \textit{profession}.
Three cities: \texttt{Beijing}, \texttt{Shanghai} and \texttt{Guangzhou} of China are chosen as \textit{location} categories, and the four most popular professions:\texttt{financial industry}, \texttt{legal profession}, \texttt{architect} and \texttt{clinical treatment} are chosen as \textit{profession} categories.

\begin{table}[t]
    \centering
    \small
    \begin{tabular}{c|c|c|c|c}
        \hline
        Dataset & DBLP & \multicolumn{3}{c}{Zhihu}\\
        \hline\hline
        Attribute  & conference & gender & location & profession \\
        \hline
        $\#Labels$ & 2          & 2      & 3        & 4          \\ \hline
        $|V_n|$ & 629,814 & \multicolumn{3}{c}{50,000}\\ \hline
        $|E_{nn}|$ & 632,751 & \multicolumn{3}{c}{241,098}\\ \hline
        $|E_{nc}|$ & 648,243 & \multicolumn{3}{c}{1,270,900}\\ \hline
    \end{tabular}
    \caption{Dataset overview.}
    \label{tab:network_overview}%
    \vspace{-0.8em}
\end{table}%

% !TEX root = main.tex

\begin{table*}[t]
    \centering
    \small
    \begin{tabular}{l|l|ccccccccc}
        \hline
        \multicolumn{2}{c|}{Algorithm}     & 10\%             & 20\%             & 30\%             & 40\%             & 50\%             & 60\%             & 70\%             & 80\%             & 90\%             \\ \hline \hline
        \multirow{3}{*}{Structure} & DW    & 87.30          & 88.27          & 88.41          & 88.88          & 88.90          & 88.78          & 88.48          & 89.17          & 89.20          \\
                                   & LINE  & 87.60          & 88.55          & 88.60          & 88.87          & 88.83          & 89.21          & 89.05          & 89.50          & 89.86          \\
                                   & W2V   & 86.79          & 87.79          & 88.36          & 88.60          & 89.03          & 89.39          & 89.01          & 89.82          & 90.25          \\ \hline
        \multirow{2}{*}{Content}   & D2V   & 83.80          & 84.24          & 84.78          & 84.89          & 84.93          & 85.71          & 85.48          & 85.50          & 86.49          \\
                                   & WAvg  & 85.67          & 87.15          & 87.88          & 87.96          & 88.25          & 88.71          & 88.53          & 89.09          & 89.52          \\ \hline
        \multirow{1}{*}{Combined}  & NC    & 88.18          & 89.39          & 89.47          & 89.88          & 90.16          & 90.60          & 90.77          & 91.40          & 91.77          \\ \hline
        \multirow{3}{*}{CENE}      & WAvg  & \textbf{90.29} & \textbf{91.23} & \textbf{91.67} & 91.47          & 91.79          & 92.24          & 92.25          & 92.26          & 92.27          \\
                                   & RNN   & 89.73          & 90.66          & 90.70          & 91.08          & 91.36          & 91.55          & 91.66          & 91.90          & 92.18          \\
                                   & BiRNN & 90.15          & 91.03          & 91.42          & \textbf{91.61} & \textbf{92.18} & \textbf{92.48} & \textbf{92.40} & \textbf{92.68} & \textbf{92.71} \\ \hline
    \end{tabular}
    \caption{Performance on DBLP. (The input matrix of DBLP for TADW is too large to be loaded into memory of our machine.)}
    \label{tab:result-dblp}
    \vspace{-0.8em}
\end{table*}

\subsection{Baseline}
% * Network embedding method: Deepwalk, LINE, word2vecf
% * Document embedding method: word2vec, doc2vec
% * Combined embedding
% To validate the performance of our approach, we compare the following methods:
We consider the following network embedding methods for experimental comparison:

\vspace{-0.5em}
\paragraph{Structure-Based Method}
\begin{itemize}
    % \vspace{-0.3em}
    \item  DeepWalk (DW)~\cite{perozzi2014deepwalk}. DeepWalk learns vertex embeddings by using the skip-gram model over vertex sequences generated through random walking on the network. %\vspace{-0.3em}
          % DeepWalk adopts random walks to generate vertex sequences, then taking these sequences as sentences the skip-gram model is used to generate vertex embeddings.
          % \footnote{\url{https://github.com/phanein/deepwalk}}
    \item  LINE~\cite{tang2015line}. LINE takes both 1-order and 2-order proximity into account and the concatenation of these two representations is used as the final embedding. %\vspace{-0.3em}
          % \footnote{\url{https://github.com/tangjianpku/LINE}}
    \item Word2vec (W2V). We include an additional baseline that uses Word2vec~\cite{mikolov2013efficient} to directly learn vertex embeddings from node-node links. Specifically, we treat each vertex $u$ as the word and all its neighbors as its context.
          Here we use the word2vecf toolkit.\footnote{\url{bitbucket.org/yoavgo/word2vecf}} %\vspace{-0.3em}
\end{itemize}

\vspace{-0.5em}
\paragraph{Content-Based Method}
\begin{itemize}
    % \vspace{-0.3em}
    \item Doc2vec (D2V)~\cite{le2014distributed}. Doc2vec is an extension of word2vec that learns document representation by predicting the surrounding words in contexts sampled from the document. Here we use the Gensim implementation\footnote{\url{radimrehurek.com/gensim/models/doc2vec.html}}. %\vspace{-0.3em}
          % Doc2vec \cite{le2014distributed} is a paragraph representation algorithm. In our experiments, we treat all text information as a document and the vector learned by doc2vec is regarded as the embedding of vertex.
    \item Word Average (WAvg). Similar to the WAvg setting in our model (CENE), we are also interested to see how well word average performs when trained separately. %\vspace{-0.3em}
          % Word embedding average is a fairly strong baseline for text classification \cite{joulin2016bag}. Just like Doe2vec, we treat all text information as a document and use the average of the embedding of each word as the content embedding.
\end{itemize}

% \vspace{-0.5em}
\paragraph{Combined Method}
\begin{itemize}
    % \vspace{-0.3em}
    \item Naive Combination (NC). We concatenate the two best-performing network embeddings learned using structure-based methods and content-based methods respectively. %\vspace{-0.3em}
    \item TADW~\cite{yang2015network}. TADW integrates content information into network embeddings by factorizing a text-associated matrix. %\vspace{-0.3em}
\end{itemize}

%\textbf{CENE:}
%\begin{itemize}
%    \item Word Average. The CENE model which uses word averaging as the embedding of contents.
%    \item RNN. The CENE method which use RNN to model content embedding.
%\end{itemize}

\subsection{Evaluation}

We evaluate our network embeddings on the node classification task.
Following the metric used in previous studies~\cite{perozzi2014deepwalk,tang2015line}, we randomly sample a portion ($T_R$, from 10\% to 90\%) of the labeled vertices as training data, with the rest of the vertices for testing. We use the scikit-learn \cite{scikit-learn} to train logistic regression classifiers. For each $T_R$, the experiments are executed independently for 40 times and we report the averaged Micro-F$_1$ measures.

\subsection{Training Protocols}
The initial learning rate is set to $\eta_0=0.025$ for CENE$_\text{WAvg}$ and $\eta_0=0.01$ for CENE$_\text{RNN}$ and CENE$_\text{BiRNN}$.
The dimension of the embeddings for both nodes and contents is set to 200.
Word embeddings are pretrained using the whole set of contents associated with the network, with dimension of 200.
In addition, the negative sampling size $SN_{nn}^u$ is 15 for all methods, and $SN_{nc}^u$ is 25 for CENE; the total number of samples $T$ is 10 billion for LINE (1st) and LINE (2nd) as shown in \newcite{tang2015line}; window size $win = 5$, walk length $t = 40$ and number of walks per vertex $\gamma =50$ for DeepWalk.

% !TEX root = main.tex

\section{Results and Analysis}
\label{sec:experiment}

\subsection{Classification tasks}

% * Result
% * Analyze

The classification results are shown in
Table \ref{tab:result-dblp} (DBLP), Table \ref{tab:result-zhihu-gender} (Zhihu-Gender),
Table \ref{tab:result-zhihu-location} (Zhihu-Location) and Table \ref{tab:result-zhihu-profession} (Zhihu-Profession).
The proposed CENE consistently and significantly outperforms both structure-based and content-based methods on all different datasets and most training rations, demonstrating the efficacy of our approach.
% Actually, for Zhihu, CENE methods perform better than these approaches when they are given 90\% of the data with only 30\% labeled data.

Besides, we have the following interesting observations:

\begin{enumerate}
    %\vspace{-0.3em}
    \item For most tasks, simple concatenation of structure-based methods and content-based methods yeilds improvements, showing the importance of both network structure and contents. %\vspace{-0.3em}
    \item Despite the simplicity, CENE$_\text{WAvg}$ obtains promising results in general, outperforming most of the baseline methods by a significant margin. Furthermore, CENE$_\text{RNN}$ and CENE$_\text{BiRNN}$ perform better than WAvg in most cases. %\vspace{-0.3em}
    \item BiRNN works better than RNN in DBLP, while RNN is better in Zhihu. The main factor here is the average sentence length in DBLP (25) and Zhihu (11). As discussed earlier (the Introduction part), BiRNN is more powerful for longer sentences. %\vspace{-0.3em}
    \item Content-based methods work generally better than structure-based methods on Zhihu, but worse on DBLP. This observation implies that structural relationships are more indicative than contents in DBLP, that is, papers tend to cite papers within the same area. Zhihu, however, is an interest-driven network, and thus contents are more important for node representation. % \vspace{-0.3em}
    \item TADW performs poorly on Zhihu. This is mainly because TADW is originally designed for networks where each node has only one document. However, nodes on Zhihu networks may follow multiple questions and the contents are relatively independent. %\vspace{-0.3em}
\end{enumerate}

% !TEX root = ../main.tex

\begin{table*}[t]
    \centering
    \small
    \begin{tabular}{l|l|ccccccccc}
        \hline
        \multicolumn{2}{c|}{Algorithm}     & 10\%             & 20\%             & 30\%             & 40\%             & 50\%             & 60\%             & 70\%             & 80\%             & 90\%             \\ \hline \hline
        \multirow{3}{*}{Structure} & DW    & 60.36          & 61.77          & 62.44          & 62.89          & 63.23          & 63.59          & 63.52          & 63.40          & 63.42          \\
                                   & LINE  & 62.03          & 63.29          & 63.81          & 64.17          & 64.41          & 64.67          & 64.89          & 64.85          & 64.76          \\
                                   & W2V   & 63.24          & 64.88          & 65.50          & 65.84          & 66.02          & 66.24          & 66.39          & 66.50          & 66.31          \\ \hline
        \multirow{2}{*}{Content}   & D2V   & 62.90          & 64.41          & 65.02          & 65.41          & 65.67          & 65.94          & 65.93          & 66.00          & 66.40          \\
                                   & WAvg  & 66.52          & 67.06          & 67.35          & 67.39          & 67.61          & 67.78          & 67.60          & 67.54          & 67.04          \\ \hline
        \multirow{2}{*}{Combined}  & NC    & 67.69          & 69.98          & 70.88          & 71.41          & 71.84          & 72.14          & 72.21          & 72.51          & 72.58          \\
                                   & TADW  & 59.51          & 59.25          & 59.42          & 59.55          & 59.52          & 59.69          & 59.57          & 59.36          & 59.54          \\ \hline
        \multirow{3}{*}{DLNE}      & WAvg  & 72.17          & 75.22          & 76.58          & 77.14          & 77.67          & 77.91          & 78.22          & 78.52          & 78.36          \\
                                   & RNN   & \textbf{75.83} & \textbf{77.54} & \textbf{78.21} & \textbf{78.39} & \textbf{78.82} & \textbf{78.95} & \textbf{79.08} & \textbf{79.35} & \textbf{79.15} \\
                                   & BiRNN & 73.95          & 76.19          & 76.94          & 77.22          & 77.59          & 77.77          & 77.75          & 77.96          & 78.09          \\ \hline
    \end{tabular}
    \caption{Performance on Zhihu-Gender.}
    \label{tab:result-zhihu-gender}
\end{table*}

\begin{table*}[t]
    \centering
    \small
    \begin{tabular}{l|l|ccccccccc}
        \hline
        \multicolumn{2}{c|}{Algorithm}     & 10\%             & 20\%             & 30\%             & 40\%             & 50\%             & 60\%             & 70\%             & 80\%             & 90\%             \\ \hline \hline
        \multirow{3}{*}{Structure} & DW    & 36.20          & 37.00          & 37.36          & 38.11          & 38.24          & 38.64          & 39.03          & 39.26          & 39.70          \\
                                   & LINE  & 37.47          & 38.04          & 38.59          & 38.83          & 39.17          & 39.08          & 39.56          & 39.48          & 40.15          \\
                                   & W2V   & 36.70          & 37.26          & 37.83          & 38.54          & 39.28          & 39.36          & 39.71          & 40.00          & 40.00          \\ \hline
        \multirow{2}{*}{Content}   & D2V   & 37.18          & 37.72          & 38.51          & 38.96          & 39.29          & 39.71          & 40.42          & 40.33          & 40.53          \\
                                   & WAvg  & 41.02          & 41.35          & 41.61          & 41.88          & 42.26          & 42.34          & 42.82          & 42.88          & 43.26          \\ \hline
        \multirow{2}{*}{Combined}  & NC    & 38.06          & 38.75          & 39.36          & 39.90          & 40.25          & 40.42          & 40.58          & 40.91          & 41.29          \\
                                   & TADW  & \textbf{43.79} & \textbf{43.83} & 43.89          & 43.89          & 43.96          & 43.86          & 43.89          & 43.99          & 44.00          \\ \hline
        \multirow{3}{*}{DLNE}      & WAvg  & 40.81          & 42.45          & 44.04          & 45.13          & 46.05          & 46.96          & 47.04          & 48.06          & 47.24          \\
                                   & RNN   & 40.88          & 43.53          & \textbf{45.45} & \textbf{46.16} & \textbf{47.05} & \textbf{47.63} & \textbf{47.92} & \textbf{48.24} & \textbf{48.37} \\
                                   & BiRNN & 39.15          & 40.98          & 42.41          & 43.46          & 43.90          & 44.58          & 44.87          & 45.39          & 45.05          \\ \hline
    \end{tabular}
    \caption{Performance on Zhihu-Location.}
    \label{tab:result-zhihu-location}
\end{table*}

\begin{table*}[!t]
    \centering
    \small
    \begin{tabular}{l|l|ccccccccc}
        \hline
        \multicolumn{2}{c|}{Algorithm}     & 10\%             & 20\%             & 30\%             & 40\%             & 50\%             & 60\%             & 70\%             & 80\%             & 90\%             \\ \hline \hline
        \multirow{3}{*}{Structure} & DW    & 47.47          & 48.26          & 49.05          & 49.45          & 50.47          & 51.09          & 51.72          & 52.93          & 51.48          \\
                                   & LINE  & 49.27          & 51.01          & 51.79          & 51.87          & 52.68          & 52.42          & 53.45          & 53.11          & 52.00          \\
                                   & W2V   & 46.08          & 48.24          & 48.91          & 50.13          & 50.16          & 50.94          & 51.24          & 52.82          & 51.26          \\ \hline
        \multirow{2}{*}{Content}   & D2V   & 44.88          & 47.07          & 48.41          & 49.67          & 50.77          & 51.41          & 51.71          & 52.56          & 53.35          \\
                                   & WAvg  & 48.31          & 51.12          & 52.94          & 54.16          & 55.37          & 55.09          & 56.22          & 56.58          & 56.80          \\ \hline
        \multirow{2}{*}{Combined}  & NC    & 53.72          & 56.70          & 58.04          & 58.96          & 60.09          & 60.47          & 61.31          & 61.68          & 61.67          \\
                                   & TADW  & 38.60          & 38.72          & 38.88          & 38.54          & 38.70          & 38.71          & 38.78          & 39.09          & 38.56          \\ \hline
        \multirow{3}{*}{CENE}      & WAvg  & \textbf{59.03} & \textbf{61.20} & \textbf{62.20} & 62.91          & 63.24          & \textbf{64.35} & 64.44          & 65.57          & 65.39          \\
                                   & RNN   & 57.90          & 60.62          & 62.10          & \textbf{63.18} & \textbf{64.17} & \textbf{64.35} & \textbf{65.56} & \textbf{65.84} & 66.31          \\
                                   & BiRNN & 57.38          & 59.43          & 60.60          & 61.47          & 62.40          & 63.51          & 64.25          & 65.26          & \textbf{66.35} \\ \hline
    \end{tabular}
    \caption{Performance on Zhihu-Profession.}
    \label{tab:result-zhihu-profession}
\end{table*}

We further conduct experiments on another DBLP \footnote{\url{github.com/shiruipan/TriDNR}} dataset used in TriDNR \cite{pan16tri} to directly compare with it.
We examine both the original semi-supervised version of TriDNR and an unsupervised version, in which the label-node relationship is discarded.
Table \ref{tab:result-with-tridnr} shows that CENE$_\text{RNN}$ and CENE$_\text{BiRNN}$ even outperform the semi-supervised TriDNR, which is really promising.

% \begin{table}[ht]
%     \centering
%     \begin{tabular}{l|llll}
%         Algoritm             & 10\%          & 30\%          & 50\%          & 70\%          \\ \hline \hline
%         TriDNR$_\text{semi}$ & 68.7          & 72.7          & 73.8          & 74.4          \\
%         TriDNR$_\text{un}$   & 64.7          & 63.5          & 64.6          & 64.2          \\ \hline
%         CENE$_\text{WAvg}$   & 71.5          & 72.7          & 72.9          & 73.1          \\
%         CENE$_\text{RNN}$    & \textbf{72.7} & \textbf{73.8} & 74.0          & 74.4          \\
%         CENE$_\text{BiRNN}$  & \textbf{72.7} & 73.7          & \textbf{75.0} & \textbf{76.3}
%     \end{tabular}
%     \caption{Performance compared with TriDNR}
%     \label{tab:result-with-tridnr}
% \end{table}

\begin{table}[h]
    \centering
    \small
    \vspace{-0.8em}
    \begin{tabular}{l|l|cccc}
        \hline
        \multicolumn{2}{c|}{Algorithm}  & 10\%           & 30\%           & 50\%           & 70\%           \\ \hline \hline
        \multirow{2}{*}{TriDNR} & Semi  & 68.7           & 72.7           & 73.8           & 74.4           \\
                                & UN    & 64.7           & 63.5           & 64.6           & 64.2           \\ \hline
        \multirow{3}{*}{CENE}   & WAvg  & 71.50          & 72.72          & 72.94          & 73.10          \\
                                & RNN   & 72.65          & \textbf{73.77} & 73.98          & 74.38          \\
                                & BiRNN & \textbf{72.75} & 73.69          & \textbf{75.01} & \textbf{76.26} \\ \hline
    \end{tabular}
    \caption{Performance compared with TriDNR.}
    \label{tab:result-with-tridnr}
    \vspace{-0.8em}
\end{table}

\begin{figure}[htbp]
    \centering
    \vspace{-1em}
    % \vspace{-0.8em}
    \begin{subfigure}[t]{0.22\textwidth}
        \includegraphics[width=\textwidth]{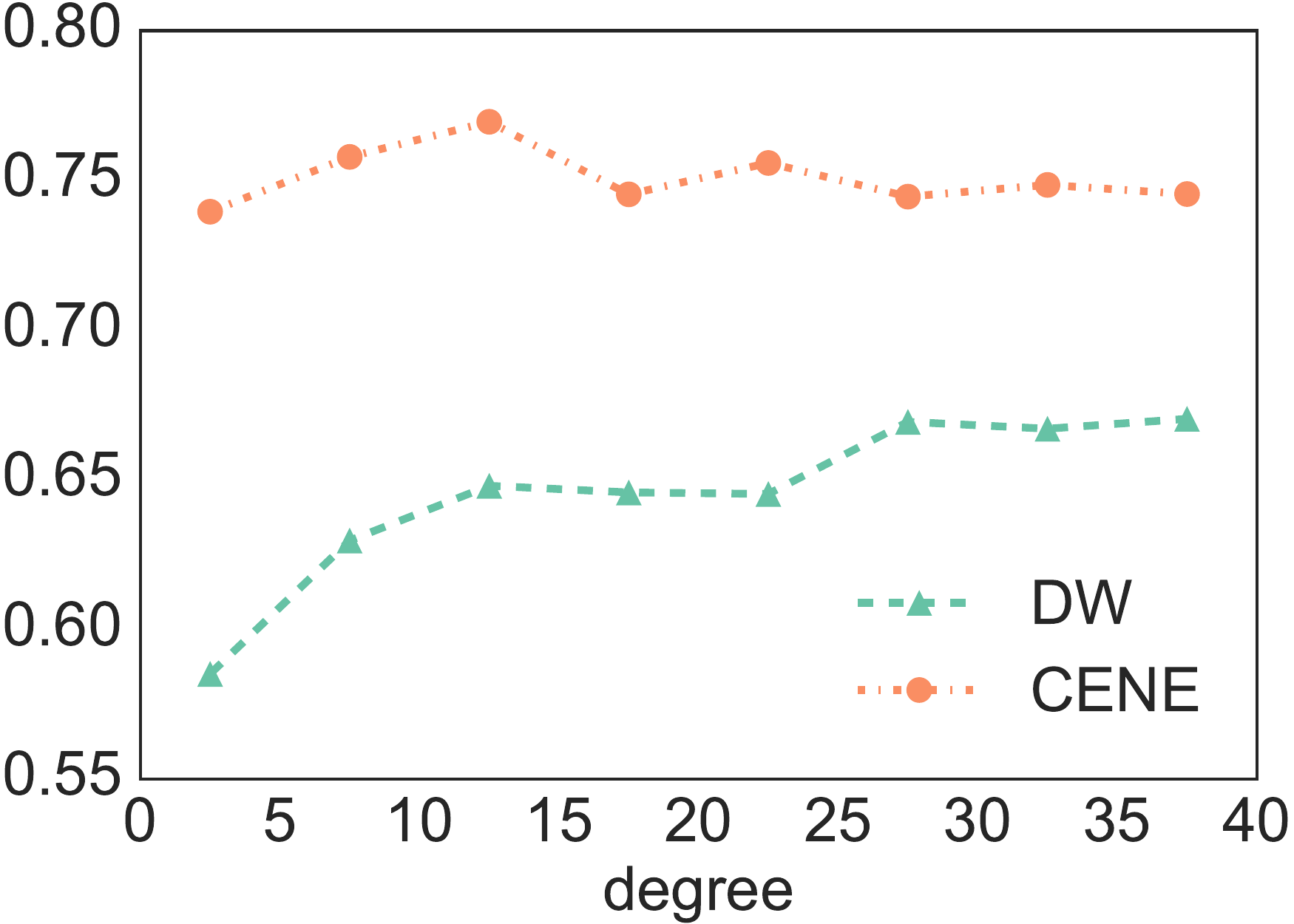}
        \caption{Performance}
        \label{fig:zhihu-gender-degree}
        \vspace{-0.8em}
    \end{subfigure}~~~~
    \begin{subfigure}[t]{0.22\textwidth}
        \includegraphics[width=\textwidth]{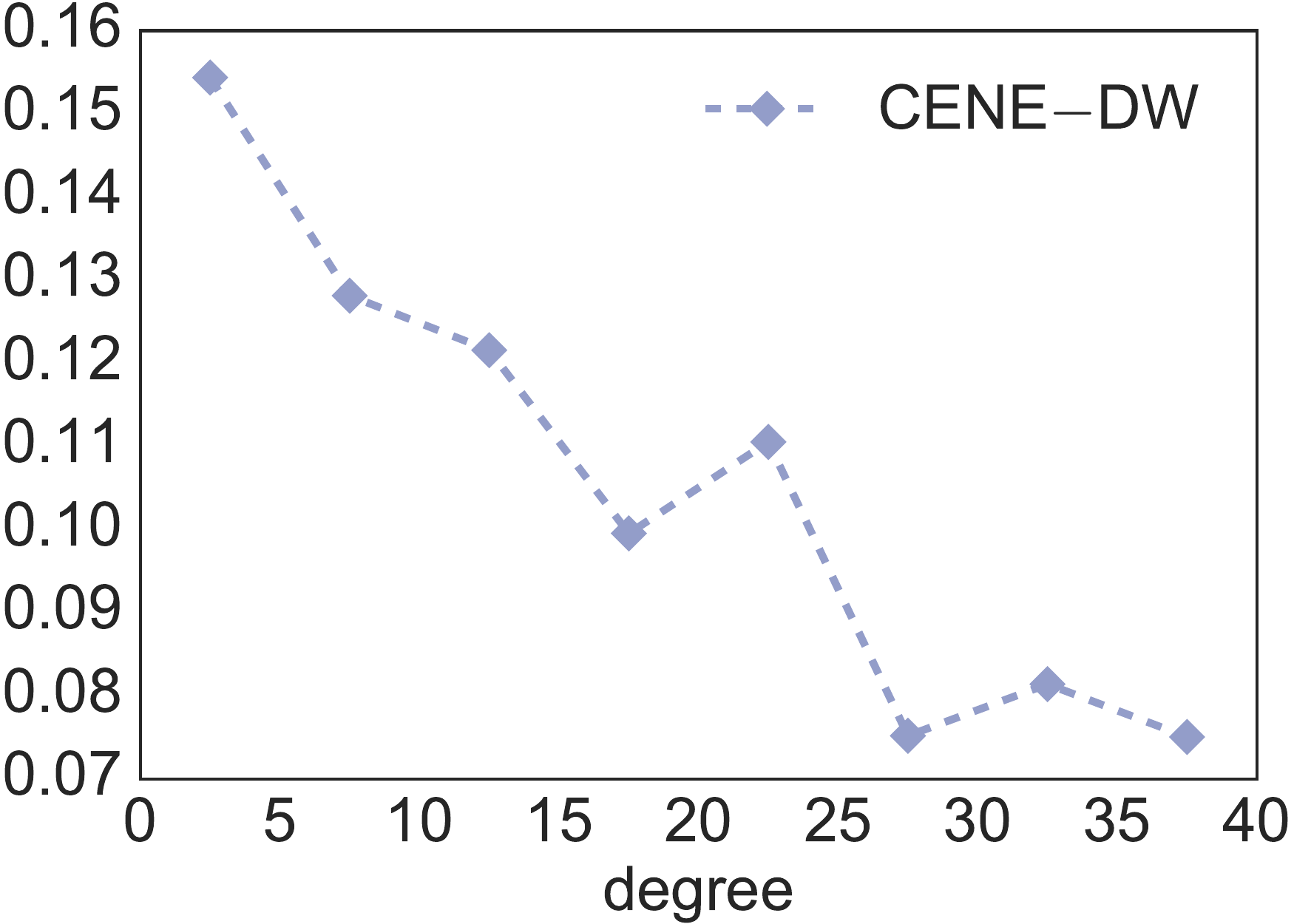}
        \caption{Performance difference}
        \label{fig:zhihu-gender-degree-diff}
        \vspace{-0.8em}
    \end{subfigure}
    \caption{Performance of Zhihu-Gender over user groups with different degrees.}
    \vspace{-1.5em}
    \label{fig:degree}
\end{figure}

% An important advantage of CENE is that it alleviates the structural sparsity in the original network.
Conventional structure-based methods perform poorly on small-degree nodes (e.g., a Zhihu user may neither follow nor be followed).
However, the introduction of content nodes would greatly alleviate the structural sparsity.
% To figure out how CENE improve the classification performance over nodes with different degrees, we also count the scores on user groups with different degrees.
Figure \ref{fig:zhihu-gender-degree} shows the classification performance of CENE over nodes with different degrees on Zhihu-Gender, compared with DeepWalk.
Figure \ref{fig:zhihu-gender-degree-diff} shows the curve of the absolute differences.
We can see CENE has a significantly larger impact on small-degree nodes, which verifies our hypothesis.
% However, with the degree number increases, it shows better performance and the gap between Deepwalk and CENE is getting smaller. That's because structure-based methods is weakly to depict the characters of nodes with few degrees, especially isolated nodes. This result suggests that CENE is able to leverage the advantages of both structure and content information.

%%%%%%%%%%%%%%%%%%%%%%%%%%%%%%%%%%%%%%%%%%%%%%%%%%%%%%%%%%%%%%%%%%%%%%%%%%%%%%%%%%%%%%%%

\subsection{Parameter Sensitivity}
% * With iteration number
% * With alpha
% * With negative sampling number

CENE has two hyperparameters: iteration number $k$ and balance weight $\alpha$. We fix the training portion to 50\% and test the classification F1 score with different $k$ and $\alpha$.

\begin{figure*}[!]
    \centering
    % \vspace{-0.2em}
    \begin{subfigure}[]{0.22\textwidth}
        \includegraphics[width=\textwidth]{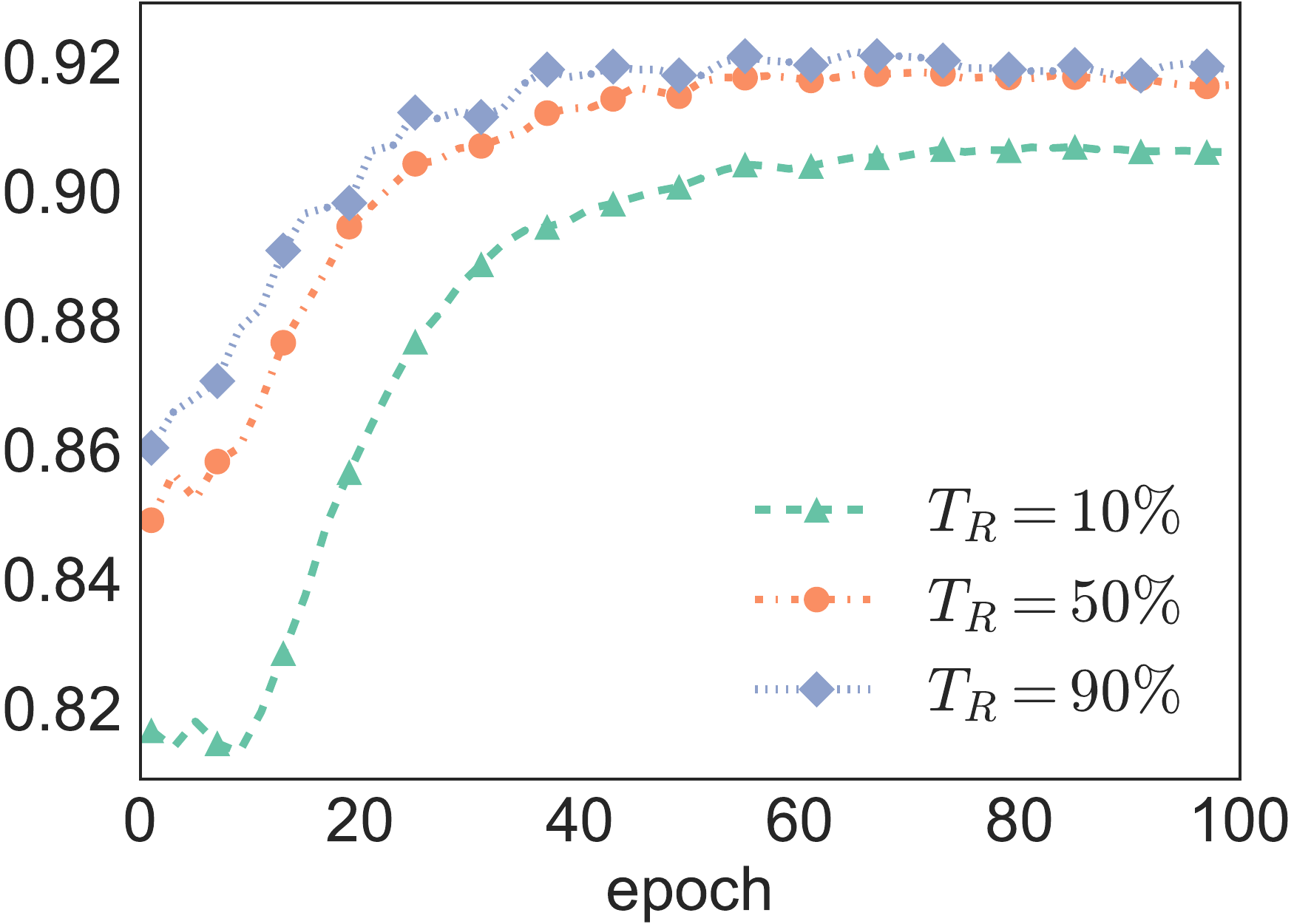}
        \caption{DBLP}
        \label{fig:dblp-iter}
        %\vspace{-0.3em}
    \end{subfigure}~~~~
    \begin{subfigure}[]{0.22\textwidth}
        \includegraphics[width=\textwidth]{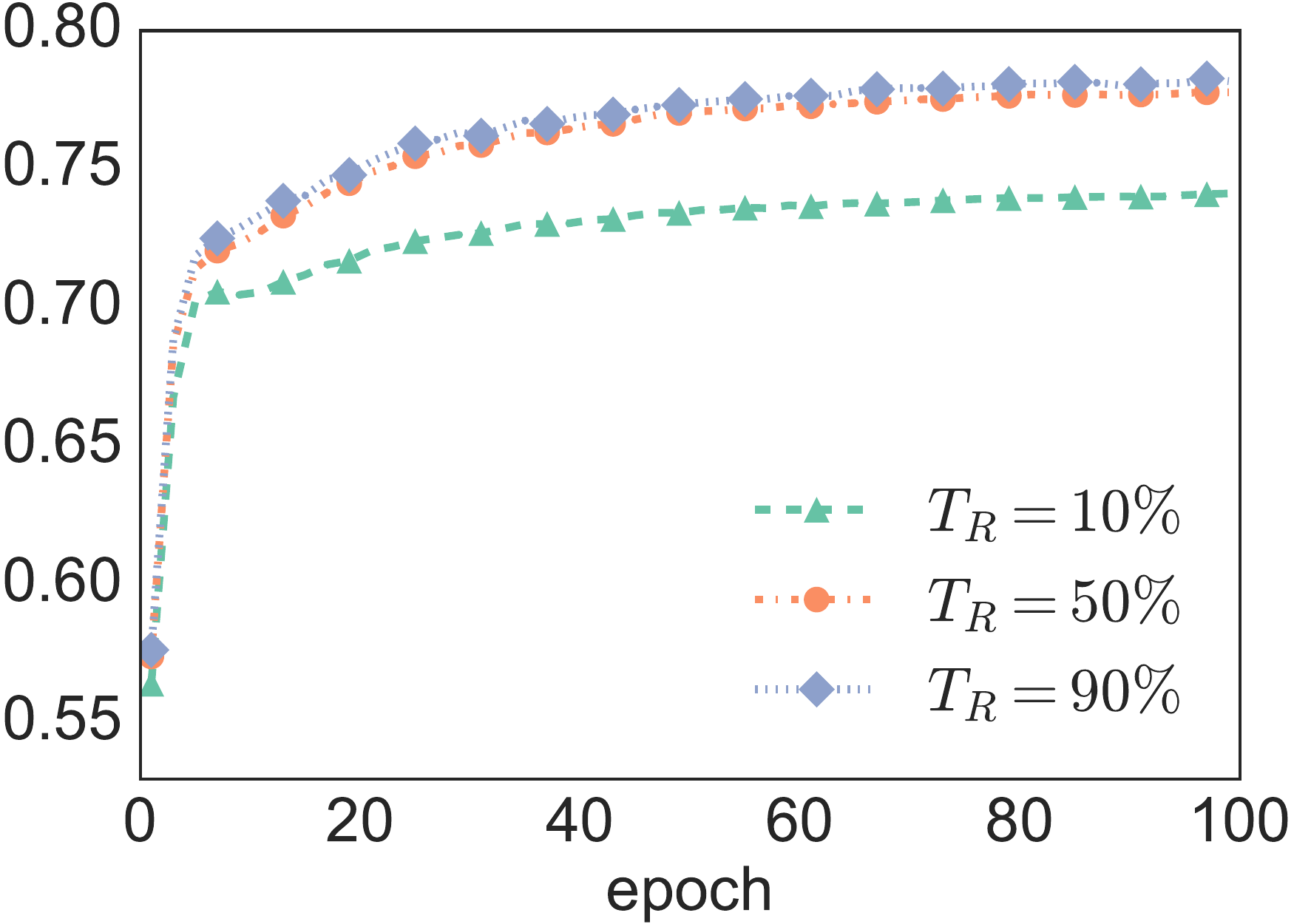}
        \caption{Zhihu-Gender}
        \label{fig:zhihu-iter-gender}
        %\vspace{-0.3em}
    \end{subfigure}~~~~
    \begin{subfigure}[]{0.22\textwidth}
        \includegraphics[width=\textwidth]{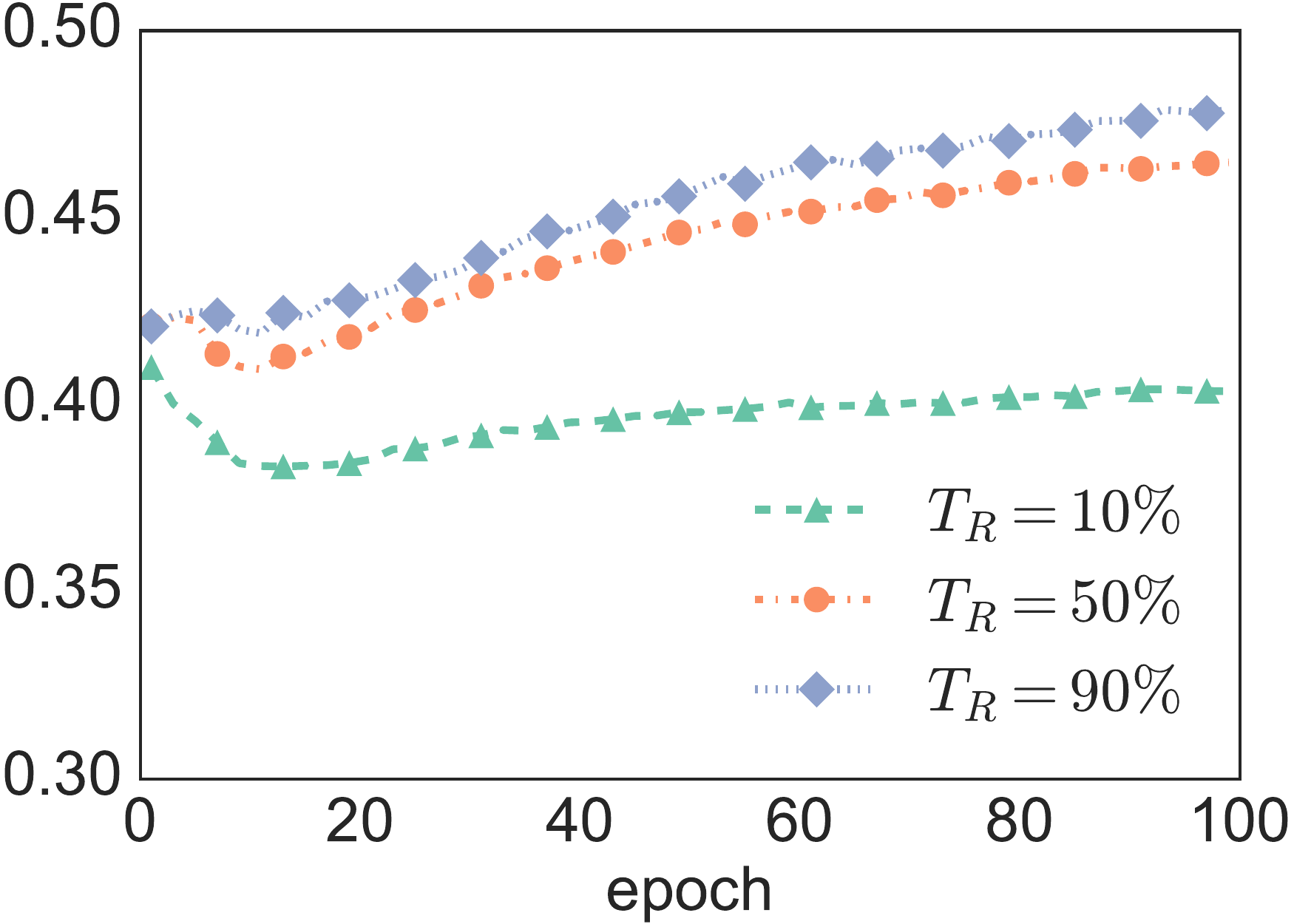}
        \caption{Zhihu-Location}
        \label{fig:zhihu-iter-location}
        %\vspace{-0.8em}
    \end{subfigure}~~~~
    \begin{subfigure}[]{0.22\textwidth}
        \includegraphics[width=\textwidth]{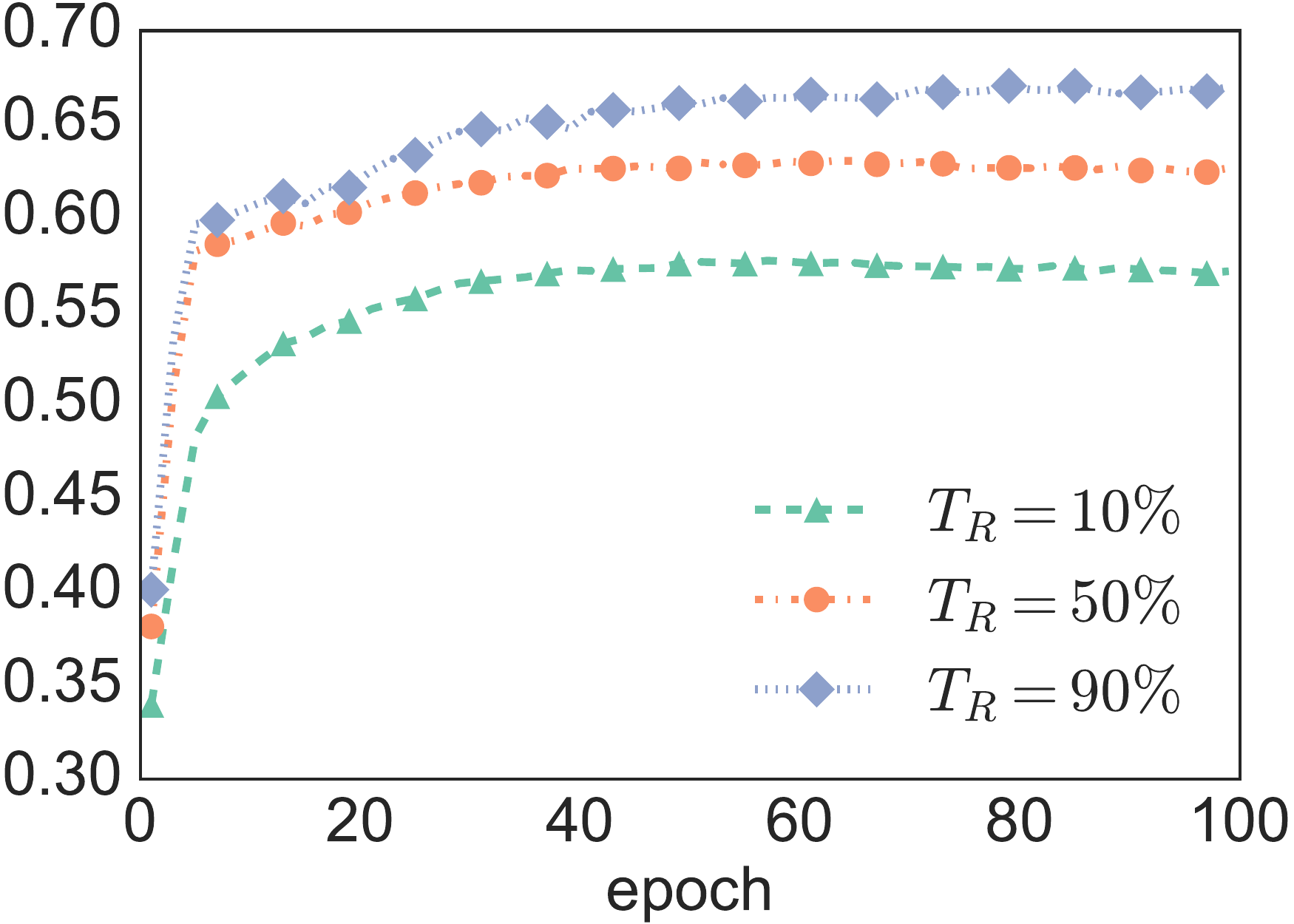}
        \caption{Zhihu-Profession}
        \label{fig:zhihu-iter-profession}
        %\vspace{-0.8em}
    \end{subfigure}
    \caption{Performance over iteration number.}
    \label{fig:iteration}
    % \vspace{-1.2em}
\end{figure*}

Figure \ref{fig:iteration} shows F1 scores with $T_R$ ranging from 10\%, 50\% to 90\%, on four different tasks. For all tasks, all of the three curves converge stably when $k$ approximates 100.

\begin{figure*}[t]
    \centering
    % \vspace{-0.5em}
    \begin{subfigure}[]{0.22\textwidth}
        \includegraphics[width=\textwidth]{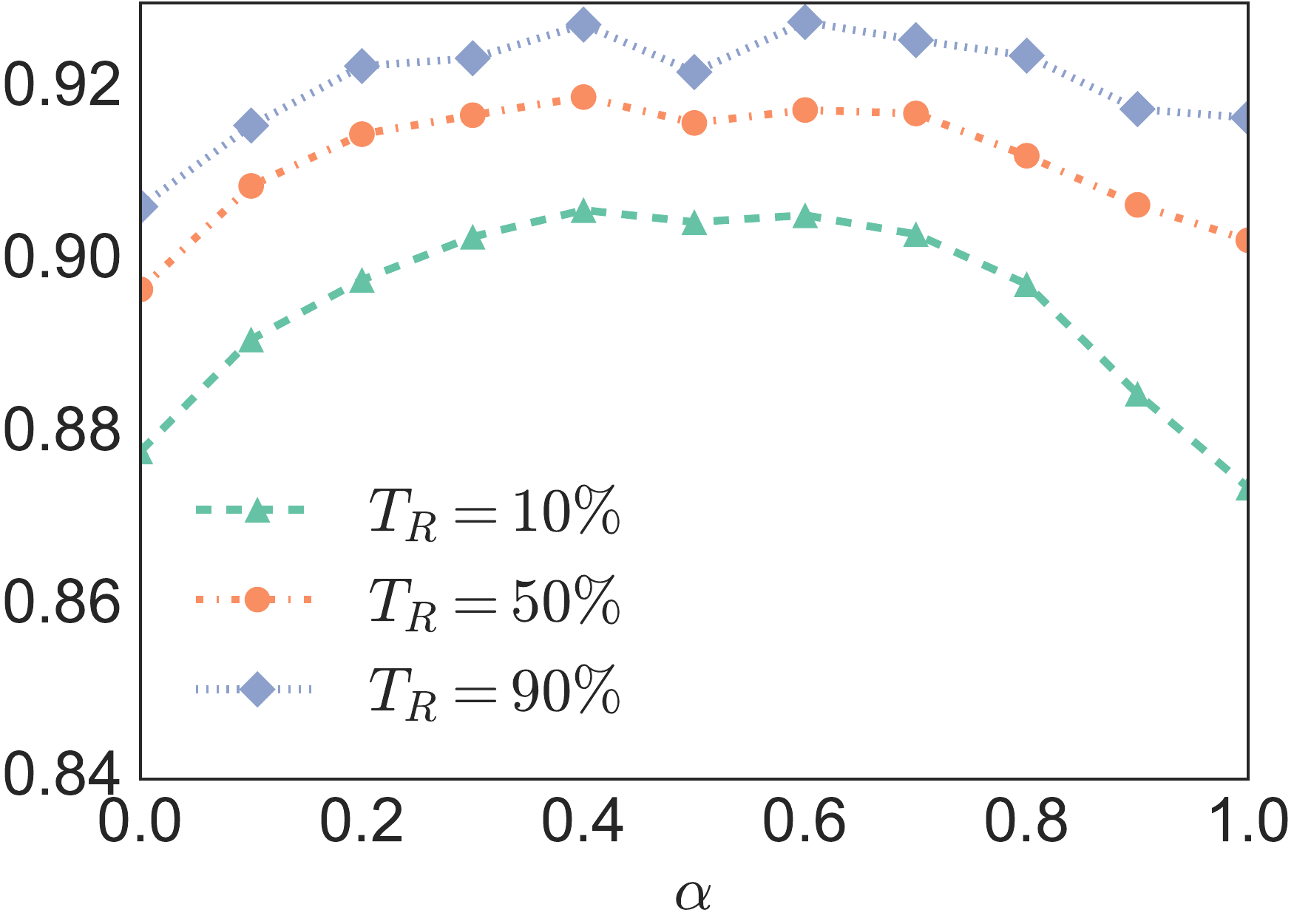}
        \caption{DBLP}
        \label{fig:dblp-alpha}
        %\vspace{-0.3em}
    \end{subfigure}~~~~
    \begin{subfigure}[]{0.22\textwidth}
        \includegraphics[width=\textwidth]{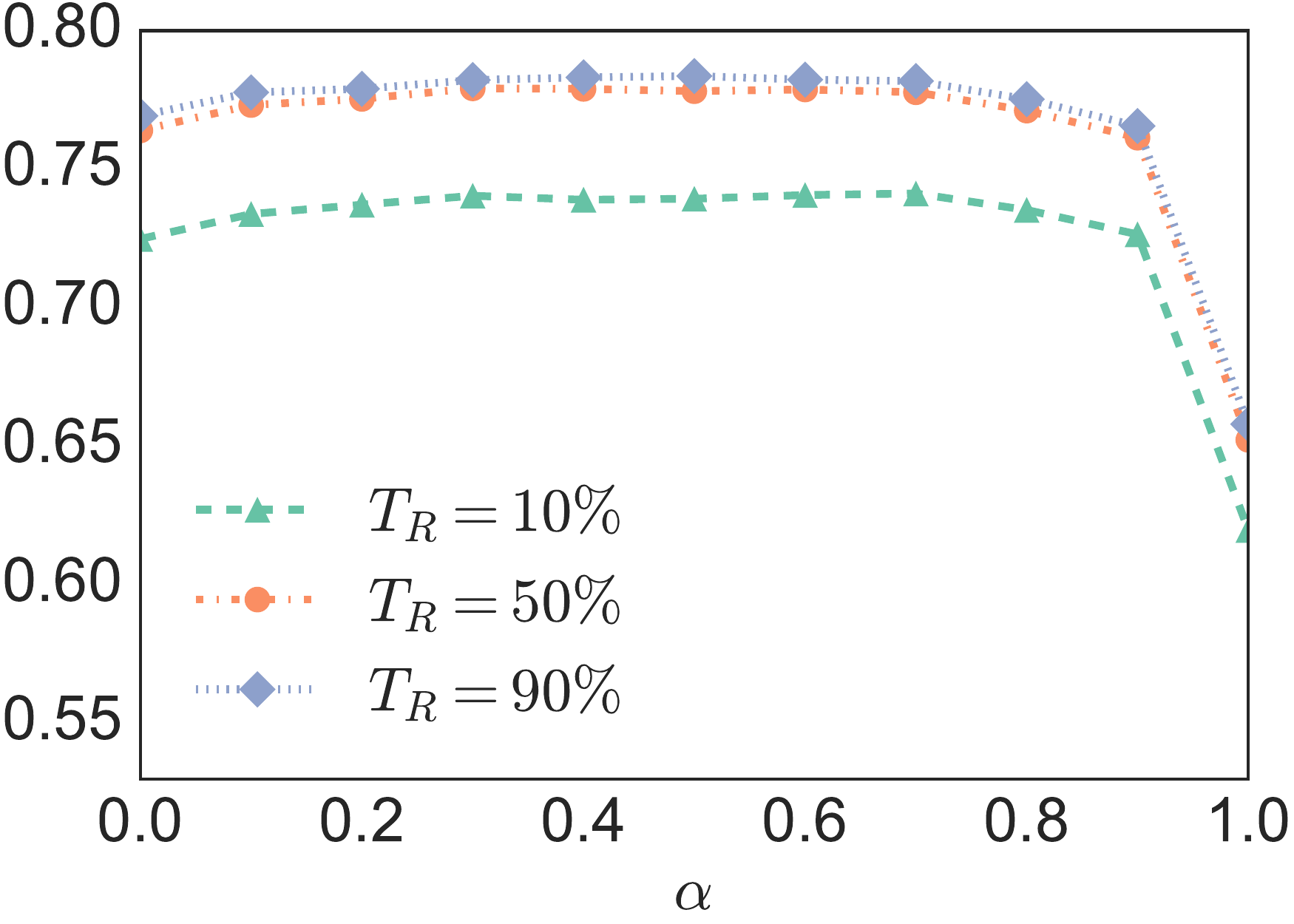}
        \caption{Zhihu-Gender}
        \label{fig:zhihu-alpha-gender}
        %\vspace{-0.3em}
    \end{subfigure}~~~~
    \begin{subfigure}[]{0.22\textwidth}
        \includegraphics[width=\textwidth]{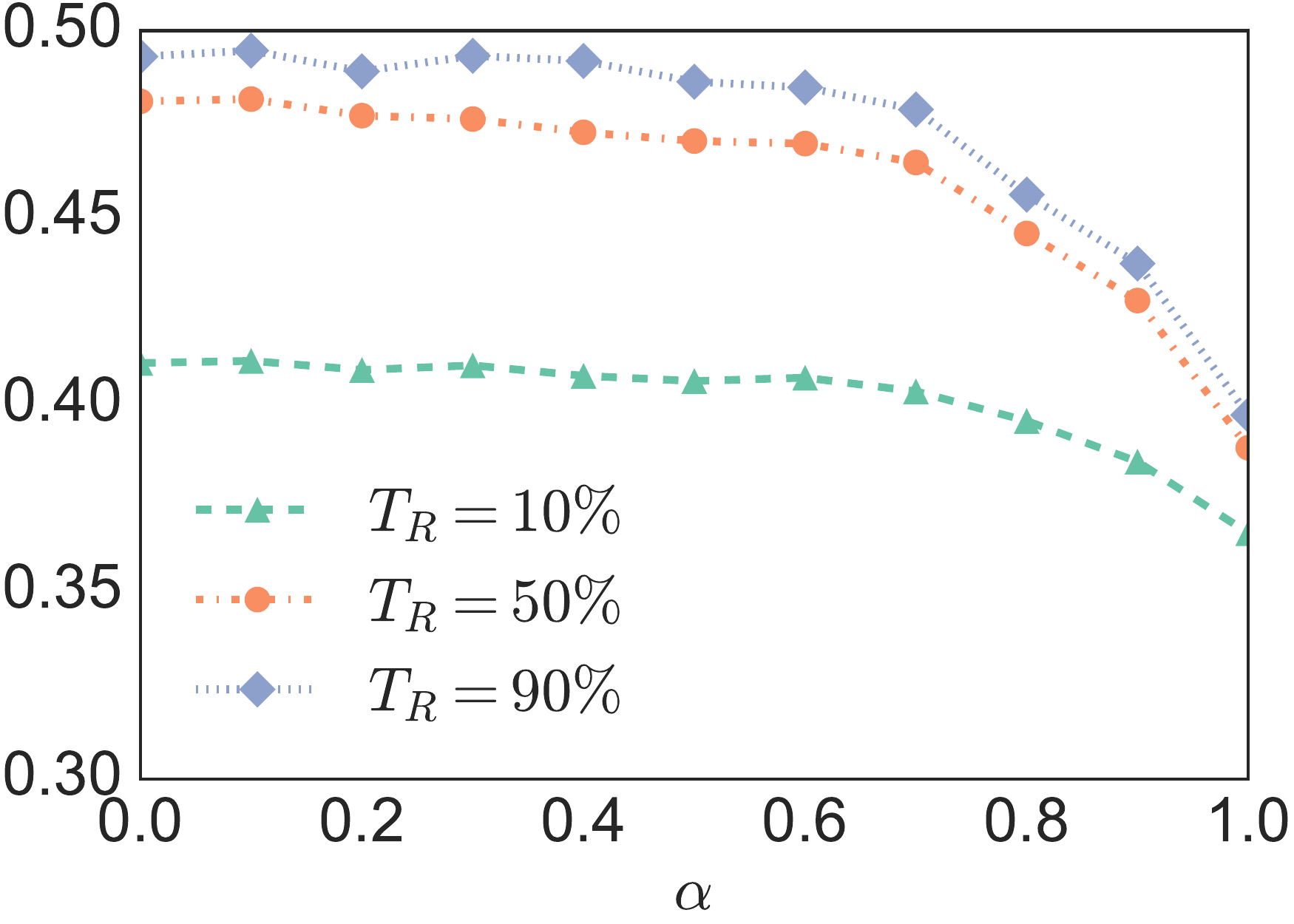}
        \caption{Zhihu-Location}
        \label{fig:zhihu-alpha-location}
        %\vspace{-0.8em}
    \end{subfigure}~~~~
    \begin{subfigure}[]{0.22\textwidth}
        \includegraphics[width=\textwidth]{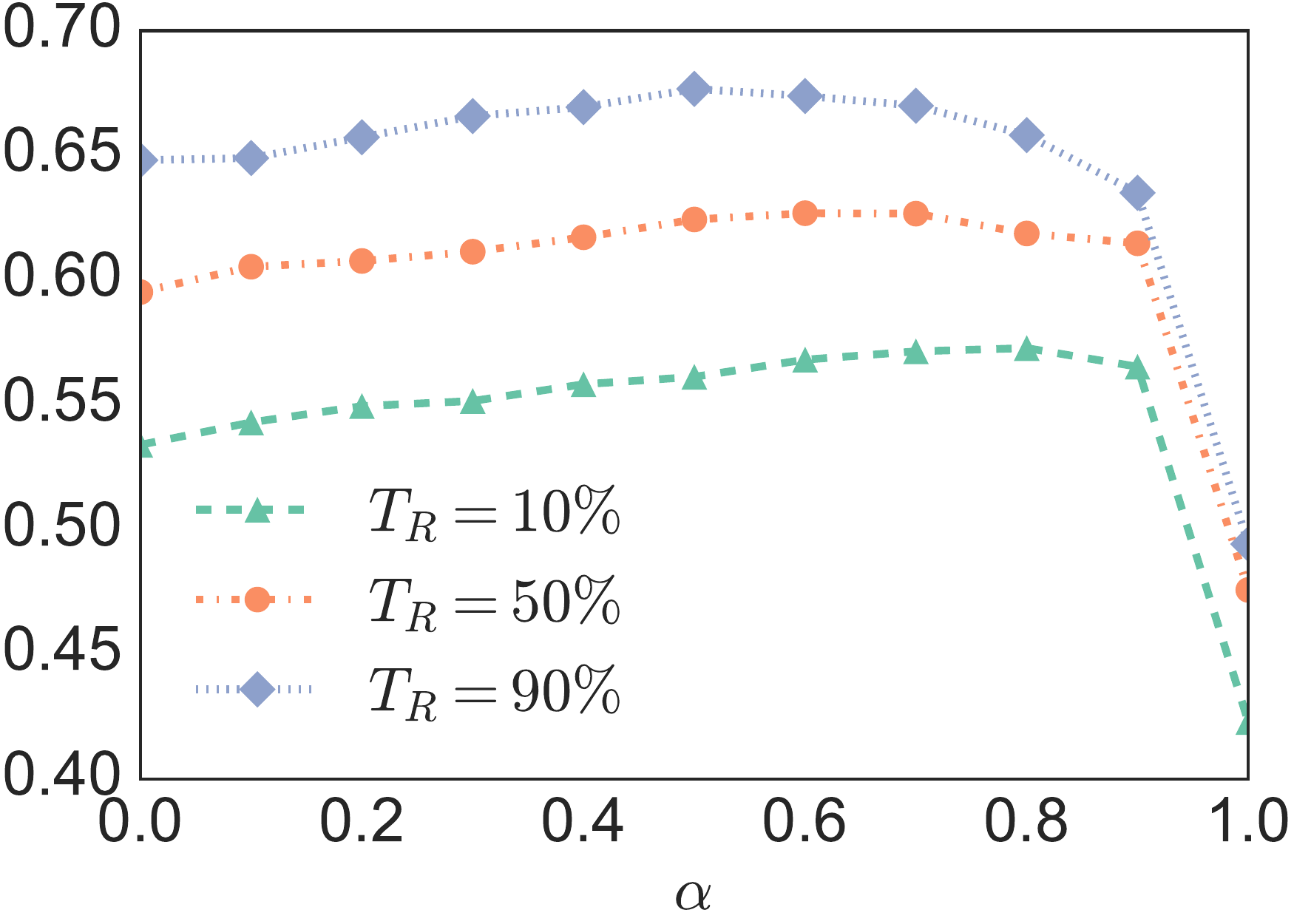}
        \caption{Zhihu-Profession}
        \label{fig:zhihu-alpha-profession}
        %\vspace{-0.8em}
    \end{subfigure}
    \caption{Performance over $\alpha$.}
    \label{fig:alpha}
    % \vspace{-1.2em}
\end{figure*}

Figure \ref{fig:alpha} shows the effect of $\alpha$. Note that if $\alpha=0$, only content information will be used, and when $\alpha=1$, our model will be degenerated into a structure-based one (W2V). With $\alpha$ increasing, the performance of CENE increases at first but decreases when $\alpha$ is big enough. There is an abrupt decrease when $\alpha$ grows from 0.9 to 1.0, indicating the importance of content information. Another notable phenomenon is that for the \textit{location} attribute on Zhihu, performance keeps dropping as $\alpha$ increases. This observation makes sense since one of the critical advantage of social networks is to break up the regional limitation, so the network structure provides little hint or even noise for identifying users' real locations.
% In this situation, larger $\alpha$ means more structural noise, which eventually reduce the final performance.

% !TEX root = main.tex

\section{Conclusion}

In this paper, we present CENE, a novel network embedding method which leverages both structure and textual content information in a network by regarding contents as a special kind of nodes. Experiments on the task of node classification with two real world datasets demonstrate the effectiveness of our model.
Three content embedding methods are investigated, and we show that deeper models (RNN and BiRNN) are more competent for text modeling. For future work, we will extend our methods to networks with more diverse contents such as images.

% !TEX root = main.tex
\section{Acknowledgments}

This work was supported by the National Basic Research Program (973 Program) of China via Grant 2014CB340503, the National Natural Science Foundation of China (NSFC) via Grant 61472107 and 61133012.

% \newpage
\bibliography{main}
\bibliographystyle{aaai}
\end{document}